\documentclass[11pt]{article}
\usepackage{graphicx}
\newcommand{\f}{\begin{equation}}
\newcommand{\ff}{\end{equation}}
\newcommand{\blankline}{\vskip .3cm}
\newcommand{\beq}{\begin{equation}}
\newcommand{\eeq}{\end{equation}}

\begin{document}
%



\title{Generic predictions of quantum theories of gravity} 

\author{Lee Smolin \thanks{Email
address:lsmolin@perimeterinstitute.ca}\\
\\
Perimeter Institute for Theoretical Physics,\\
Waterloo, Ontario N2J 2W9, Canada, and \\
Department of Physics, University of Waterloo,\\
Waterloo, Ontario N2L 3G1, Canada\\
\\
}

\maketitle

\begin{abstract}

I discuss generic consequences (sometimes called ``soft predictions")  of a class
of background independent quantum theories of spacetime called {\it causal spin network theories}. These are theories whose
kinematics and dynamics is based on the evolution of labeled graphs, by local moves, such as in loop quantum gravity and spin foam models.   Some generic consequences are well known, including
the discreteness of quantum geometry,  the elimination of spacetime singularities, the  entropy of black hole and cosmological horizons and the fact that positive cosmological constant spacetimes are hot.  Within the last few years three possible generic consequences have come to light. These are  1)   Deformed special relativity as the symmetry of the ground state, 2)  Elementary particles as  coherent excitations of quantum geometry,  3)  Locality is disordered.   I discuss some possible experimental consequences of each.  

\blankline
\blankline
{\it For inclusion in "Approaches to Quantum Gravity - toward a new understanding of space, time, and matter", edited by D. Oriti, to be published by Cambridge University Press.}

 \end{abstract}
 
\newpage
\tableofcontents

\section{Introduction}

How does a proposal for unification go from an interesting body of mathematical results to a plausible explanation of natural phenomena?  While evidence of mathematical consistency is ultimately important,  what is often decisive is that a proposed unification leads to predictions of phenomena that are both new and generic.  By generic I mean that the new phenomena are general consequences of the proposed unification and thus hold for a wide range of parameters as well as for generic initial conditions. The proposal becomes an explanation when some of those new generic
phenonena are observed.  

 Generic consequences of unification often involve processes in which the things unified transform into each other.  For example, electromagnetic waves are a generic consequence of unifying electricity and magnetism,  weak vector bosons are a generic consequence of unifying the weak and electromagnetic interactions, and light bending is a generic consequence of the equivalence principle which unifies gravity and inertia.   

Looking at history, we see that the reasons why proposals for unification succeed or fail often have to do with their generic consequences.  In successful cases the consequences do not conflict with previous experiments but are easily confirmed when looked for in new experiments. These are cases in which we come to {\it celebrate the unification.}  In bad cases the consequences generically disagree with experiment. Some of these cases still survive for some time because the theory has parameters that can be tuned to hide the consequences of the unification. But these then succumb to lack of predictability which follows from the same flexibility that allows the generic consequences to be hidden.    

It is often the case that heuristic arguments are sufficient to uncover generic consequences  of new  theories even before precise predictions can be made. It was understood that QED would lead to a Lamb shift before there were precise predictions by Feynman and  Schwinger. Einstein was able to predict that a theory based on the equivalence principle would lead to light bending before GR had been precisely formulated. Thus, uncovering generic consequences  gives both experimentalists and theorists something to focus their attention on. 

Moreover,  physicists often have  not needed to solve a theory exactly, or rigorously prove its consistency, to work out generic consequences and extract precise predictions that could be tested experimentally.  This was certainly the case with both GR and QED.  It is then incumbent on us to look at generic consequences of different proposed  unifications of quantum theory, spacetime and particle physics and try to determine if they are cases in which there is a chance to celebrate, rather than hide, their consequences. 

In this contribution I will attempt to do this for a large class of quantum gravity theories. These are theories which are {\it background independent} in that  classical fields, such as a background metric,  play no role in their formulation. To make the discussion concrete I will be interested in a large class of  theories  which I  call {\it causal spin network theories\cite{fotini-dual}.} These include the different versions of loop quantum gravity\cite{lqg, invitation} and spin foam models\cite{SF}. They include also a large class of theories describable in the general mathematical and conceptual language of LQG that have not however been derived from the quantization of any classical theory.   These theories have been much studied in the  20  years since Ashtekar wrote down his reformulation of general relativity as a gauge theory\cite{abhay}.  
There remain significant open problems; nevertheless, I hope to convince the reader that we know enough about these theories to argue for several generic consequences.   

 My intention here is to explain the basic physical reasons for these generic consequences. Consequently, the discussion will be heuristic and I will often sketch arguments that are made fully elsewhere\cite{lqg,invitation,rigorous}. 

In the next section I will list the main postulates of causal spin network theories.
Following that,  I will discuss  seven generic
consequences:

\begin{enumerate}

\item{}Discreteness of quantum geometry and ultraviolet finiteness

\item{}Elimination of spacetime singularities

\item{}Entropy of black hole and cosmological horizons

\item{}Positive cosmological constant spacetimes are hot.

\item{}Deformed special relativity

\item{}The emergence of matter from quantum geometry

\item{}Disordered locality

\end{enumerate}

The first four are well established. The next is  the subject of  recent progress and the last two are new.

 \section{Assumptions of background independent theories}
  
Four generic assumptions define a class of 
 background independent quantum gravity theories that have been the subject of most study.

\begin{itemize}

\item{}{\bf Quantum mechanics}  We assume the basic postulates of quantum mechanics.

 \item{}{\bf Partial background independence}  The theory  is formulated without reference to any fixed spacetime metric or other classical fields.  There may however be some fixed structures including  dimension, topology and boundary conditions.   General relativity is a partly background independent theory. There is an argument, to be found in 
\cite{backinv},  that any quantum theory of gravity must be so\footnote{One may also
try to make theories that are more fully background independent,  but they will not be
discussed here.}.

 \item{}{\bf Discreteness  } The Hilbert space  $\cal H$ has a countable basis given by  discrete or combinatorial structures.  The dynamics is generated by moves local in the topology of these structures. These define the events of the theory. The dynamics is specified by giving the amplitudes for the possible events. 
 
 \item{}{\bf Causality}  The histories of the theory have causal structure, in the sense that the events define a partially ordered, or causal set. 
 
 \end{itemize}
 
 There are a number of such theories, which depend on different choices for the combinatorial
 structure used to model quantum geometry.   These include dynamical triangulations\cite{CDT}, 
 causal set models\cite{CS}, quantum causal history models\cite{f-this} 
 and consistent discretization models\cite{CD}.   Important things have been learned
 from each of them. Here  I will discuss here  the following class of theories, which I  call  {\it causal spin network theories\cite{fotini-dual}}.
 
\begin{figure}[htbp]
   \centerline{\includegraphics[height=2cm]{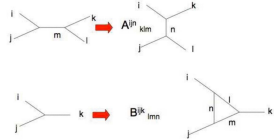}}
     \caption{The  basic local moves on trivalent graphs}
   \label{fig:localmoves3}
\end{figure}

\begin{figure}[htbp]
   \centerline{\includegraphics[height=2cm]{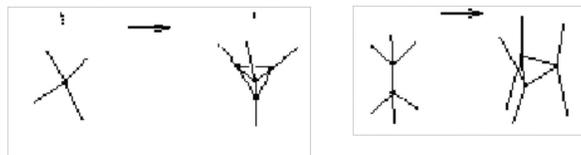}}
     \caption{The  basic local moves on four-valent graphs}
   \label{fig:localmoves4}
\end{figure}

 \begin{enumerate}
 
 \item The Hilbert space has a countable basis indexed by all embeddings, up to topology, of a
 class of  graphs $\Gamma$  in a fixed  { topological manifold}  $\Sigma$.  
 
 \item The graphs may be labeled.  If so, the labeling is determined by a choice of  a Lie algebra or quantum group $\cal A$.  The  edges of $\Gamma$ are labeled with irreducible representations $j$ of $\cal A$ and the nodes are labeled with invariants in the product of the incident representations.  Labeled graphs are  called {\it spin networks.}
 
In the nicest examples $\cal A$ is a compact Lie algebra, or its quantum deformation at a root of unit, so that the labels form a discrete set.  
 
 \item  There are a small number of local moves, for example 
 those in Figures~\ref{fig:localmoves3} and \ref{fig:localmoves4}. The amplitude of a local move is a function of the labels involved.  There are three basic kinds of moves. Expansion moves when a node is blown up to a symplex, for example a triangle, contraction moves, which are the reverse and exchange moves whereby two neighboring nodes exchange connections to other nodes.   
 
 \item A history is made of a sequence of local moves, which take the state from an initial spin network state to a final one. The moves have a partial order structure defined
 by domains of influence\cite{fotini-dual}.

 \end{enumerate}
 \begin{figure}[htbp]
   \centerline{\includegraphics[height=2cm]{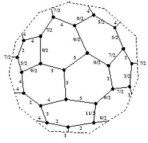}}
     \caption{An example of a spin network (from J. Baez.)}
   \label{spinnets}
\end{figure}

 We call the set of graphs, embedded in $\Sigma$ up to topology,  the fundamental configuration space,  ${\cal S}_{\Sigma, {\cal A}}$.  In the quantum theory each labeled graph embedding corresponds to an element of an orthonormal basis of the Hilbert space ${\cal H}_{\Sigma, {\cal A}}$.
 
Some theories of this kind  can be derived from classical theories which are diffeomorphism invariant gauge theories. It is the great discovery of Ashtekar that general relativity is a  theory  of this type\cite{abhay}.  This is true for any dimension and it is also true for any version of supergravity or coupling to any matter.  
 
The classical configuration space,
$\cal C$,  is then the space of connections valued in $\cal A$, on the spatial manifold $\Sigma$,  modulo gauge transformations.   The conjugate electric field turns out to be related to the  metric. There is also a diffeomorphism invariant configuration space, ${\cal C}^{diffeo}$ consisting of the orbits of   $\cal C$ under $Diff(\Sigma )$.
 
The relationship to the previous definitions  is based on the 
following two principles\cite{invitation}.

 \begin{itemize}
 
 \item{}{\bf Gauge-graph duality} The Hilbert space  ${\cal H}_{\Sigma, {\cal A}}$ is the quantization of the classical configuration space ${\cal C}^{diffeo}$ just defined.

\item{}{\bf Constrained or perturbed topological field theory } Gravitational theories, including general relativity, and supergravity can be expressed simply  in terms of constraining or perturbing topological field theories. 

\end{itemize}

 We discuss each in turn.  To realize the graph-gauge duality, we express the theory, not in terms of the connection, $A$, but in terms of the holonomy,  
 \f
 U[\gamma , A] = Pe^{\int_\gamma A }  .
 \label{wilsonloop}
 \ff
Then,  $T[\gamma ] = Tr U[\gamma , A]$ is called
the Wilson loop observable. The conjugate operator is the electric flux through a surface $S$,
 \f
 E(S,f) = \int_S E_i f_i
 \ff
 This depends also on a Lie algebra valued function on $S$, given by $f_i$. 
 These satisfy a closed Poisson algebra,
 \f
 \{  U[\gamma , A] , E(S,f) \} = l_{Pl}^2 \  Int [\gamma , S] \ U[\gamma_S , A] f
 \label{loop-surface}
 \ff
 where $ Int [\gamma , S] $ is the intersection number of the surface and loop and
 $\gamma_S$ is the loop beginning and ending at the point it intersects the surface. 
 
 Wilson loops can be extended to spin networks in the following way: to each edge 
 of a spin network $\Gamma$, write
 the holonomy in the representation indicated by the label on the edge; then tie these up with the invariants on the nodes to get a gauge invariant functional of $\Gamma$ and the 
 connection called $T[\Gamma , A]$. 
 
 There are several key features of the quantization in terms of these variables:
 
 \begin{itemize}
 
 \item{}  The Fock space plays no role at all, as that depends on a background metric. 
 
 \item  Instead, there is a uniqueness theorem\cite{unique1,unique2} that tells us that there is a unique representation of the 
algebra (\ref{loop-surface}) such that i) the Wilson loop operators create normalizable states and ii) it carries a unitary representation of the diffeomorphism group of
$\Sigma$. We call this  ${\cal H}^{kin}$ for kinematical Hilbert space. 
 
 Any consequence of  this unique representation is then a generic consequence of
 a large class of quantum gravity theories. 
 
 In  ${\cal H}^{kin}$ there is no operator that represents $A_a$, all that is represented
 are  the Wilson loops. Similarly there is no representation of infinitesimal diffeomorphisms, only finite ones. 
 
 The graph-gauge field duality is represented explicitly by a functional transform,
 \f
 \Psi [\Gamma ] = \int d\mu_{AL} (A) T[\Gamma , A] \Psi [A] 
 \ff
 where $d\mu_{AL}$ stands for the rigorously defined, {\it Ashtekar-Lewandowski measure\cite{rigorous}.} 
 
  There is a basis element for every distinct embedding of a 
 spin network ,  so ${\cal H}^{kin}$ is not separable. This is remedied by going to the subspace of diffeomorphism invariant states. 
 
 \item  The diffeomorphism invariant Hilbert space, ${\cal H}^{diffeo}$ is constructed
 by moding out by the action of $Diff( \Sigma )$ in the dual of ${\cal H}^{kin}$. There is a basis element for each (piecewise smooth) diffeomorphism class of graphs, so it is separable.   As it is constructed from a unique kinematical space by a unique operation, 
${\cal H}^{diffeo}$ is also unique. 
 
 \end{itemize}
 
 Thus,  we arrive uniquely at the kinematical structure of an  evolving spin network theory, 
 because\footnote{Because equivalence of graph embeddings under piecewise smooth embeddings is
 equivalent to topological equivalence.}
 ${\cal H}^{diffeo} = {\cal H}_{\Sigma, {\cal A}}$.  
 
 All known classical gravity theories such as GR and supergravity in any dimension are diffeomorphism invariant gauge theories. Hence they all provides examples of  causal spin network theories.  
 
But its even better than this, because the dynamics turns out to act simply on the spin network states, through local moves of the kind descrived above. This is a consequence of the second principle, which is that the dynamics of all known classical relativistic gravitational theories are arrived 
by perturbing around\cite{artem-laurent} or 
constraining topological field theories\cite{CTFT}. 

In $4$ dimensions one route to this is  through the Plebanski action\cite{plebanski}.  Pick $G= SU(2)$
and consider the action
\f
S^{BF}= \int \left ( B_{i} \wedge F^{i}  - \frac{\Lambda}{2}   B^i \wedge B_i \right )
\ff
where $B_{i}$ is a two form valued in the Lie algebra of $SU(2)$ and $F^i$ is the 
$SU(2)$ field strength.  This has no local
degrees of freedom as the field equations are
\f
F^{i} =\Lambda B^i  \ \ \ \ ; \ \ \ \ \ \  D \wedge B^{IJ}=0
\label{dS}
\ff
Now consider the following action, which differs from it by just a constraint.
\f
S^{BF}= \int \left ( B_{i} \wedge F^{i}  - \frac{\Lambda}{2}   B^i \wedge B_i 
+ \Phi_{ij}  B^i \wedge B^j  \right  )
\ff
It is not hard to see that this is an action for general relativity\cite{cosconst}. 

Starting with the action in this form, one can write a path integral representation of the dynamics 
of the spin network embeddings\cite{lqg,SF,CTFT}.  
Details are discussed elsewhere in this volume, the result is 
to give amplitudes to a set of local  moves.

 \section{Well studied generic consequences}
 
Let us begin with some well studied  generic consequences of the class of theories we have just described.  
 
 \subsection{Discreteness of quantum geometry and ultraviolet  finiteness}
 
It is well understood that such theories are generically ultraviolet finite.  The demonstrations of finiteness 
depend only on the assumptions that lead to the unique ${\cal H}^{diffeo}$ and they are now confirmed by rigorous 
results\cite{rigorous}. But the reason these theories are discrete and finite can also be understood  intuitively.  
The key point is that Wilson loop operators create normalizable states. This means that they realize precisely an old conjecture about quantum non-abelian gauge theories which is that the electric flux is quantized so the operators that measure total electric flux through surfaces have discrete spectra.  This used to be called the dual superconductor
hypothesis.  This is relevant for quantum gravity because {\it the uniqueness theorem tells us that the Hilbert space of any quantum theory of gravity describes a dual superconductor: the graphs are then the states of quantized electric flux. }

In the connection to gravitational theories the total electric flux through a surface translates to the area of the surface.  Hence the areas of all surfaces are quantized, and there is a smallest non-zero area eigenvalue. This turns out to extend to other geometrical observables including volumes, angles, and lengths. 
 
This discreteness of quantum geometry in turn implies that the theory is ultraviolet finite. The theory has no states in which areas, volumes or lengths smaller than Planck scale are meaningfully defined.  There are consequently no modes with wavelength smaller than the Planck length. It has also been shown that for a class of theories the path integral is ultraviolet finite\cite{SF}.

It can be asked whether the volume or area operators are physical observables,  so that their discreteness is a physical prediction. The answer is yes. To show this  one may first gauge fie the time coordinate, to give the theory in a version where there is an Hamiltonian evolution operator. Then one can construct the diffeomorphism invariant operator representing the volume of the gauge fixed spatial slices\cite{fixed}. In cases where one fixes a spatial boundary, the area of the boundary is also a physically meaningful operator. One can also define diffeomorphism invariant area operators by using physical degrees of freedom to pick out the surfaces to be measured\cite{diffeoinv}.  In several cases,  such physically defined geometrical observables have been shown to have 
the same discrete spectra as their kinematical counterparts\cite{diffeoinv}. Hence, the discreteness is a true generic consequence of these theories. 

\subsection{Elimination of spacetime singularities}
  
There has long been the expectation that spacetime singularities would be eliminated in quantum theories of gravity. In the context of LQG-like models this has been investigated so far in the context of a series of models\cite{bojowald}-\cite{oliver-viqar}. 
These have a reduced
number of quantum degrees of freedom corresponding to approximating the spacetime near a spacetime or black hole singularity by its homogeneous degrees of freedom. In the cosmological case the number of degrees of freedom is finite\cite{bojowald}, while in the black hole case the theory is a $1+1$ dimensional field theory, which is the symmetry of the 
interior of a Schwarzchild black hole\cite{oliver-viqar}.

All results so far confirm the expectation that the spacelike singularities are 
removed and replaced by bounces\cite{bojowald}-\cite{oliver-viqar}.  Time continues to the future of where the singularity would have been and the region to the future is expanding. 

These models are  different from the old fashion quantum cosmological models, based on ${\cal L}^2 (R^+ )$.  The key features by which they differ parallel the features mentioned above of the full diffeomorphism invariant quantum field theories. 
For example,  as in the full field theory, 
there is no operator corresponding to $A_a^i$, rather the connection degrees of 
freedom (which are the variables conjugate to the spatial metric) are represented
by the exponential (\ref{wilsonloop}).  
The elimination of singularities can be directly tied to the features of this new quantization. Thus, there is reason to expect that the same discreteness will apply to the full diffeomorphism invariant quantum field theory. Work aimed at resolving this question is in progress. 

\subsection{Entropy of black hole and cosmological horizons}
  
Generic LQG theories have a universal mechanism for describing states on certain kinds of boundaries, which includes black hole and cosmological horizons.  
In the presence of a boundary, we have
to add a boundary term to get a good variational principle.
The details are
described in \cite{linking} the key point is that through the connection to topological quantum field theory the boundary theories end up described in terms of a $2+1$ dimensional topological field theory, which is Chern-Simons theory. This follows from the fact that the  deSitter or AdS spacetime represent solutions to the pure topological field theory, this implies that the topological field theory should dominate on the boundary of an 
asymptotically dS or AdS spacetime\cite{linking,cosconst}. It turns out that the same conditions hold
on horizons\cite{kirill,isolated}. In all these cases, the 
boundary term is of the form, 
\f
S^{boundary}= \frac{k}{4\pi} \int Y^{CS} (A)
\ff
where $A$ is the pull back of a connection one form to the boundary.  The connection with Chern-Simons theory is a direct consequence of the relationship of general relativity to  topological field theory and hence is generic.   

Chern-Simons theory is used to describe  anyons in $2+1$ dimensional condensed matter physics.   The states are labeled by punctures
on the two dimensional sphere which is the spatial cross-section of the horizon. 
The punctures are points where the graphs attach to the boundary, and serve also as quanta of area on the boundary. As a result of the boundary conditions that identify the surface as a horizon, the connection is constrained to be flat everywhere except at the punctures.  The physics on a horizon is then identical to that of a system of anyons, with the area being proportional to the total charge carried by the anyons. 

Physicists know how to count the states of such $2+1$ dimensional theories. Not surprisingly, the entropy ends up proportional to the area. Getting the constant of proportionality right requires fixing a 
constant, the Immirzi constant\footnote{A heuristic argument that fixes the value of the constant in terms of
a correspondence with the quasi normal mode spectrum was given by Dreyer\cite{olaf}. When the states
of the horizon are correctly counted, one gets the same 
value\cite{symmetry-entropy,countings}. }. Once that is done all results, for all black hole and cosmological horizons, agree with Hawking's prediction, to leading order\cite{olaf,symmetry-entropy}. Past leading order there are corrections to the black hole entropy and thermal spectrum which are quantum gravity effects\cite{bh-corrections}.  These corrections introduce a fine structure into the Hawking radiation, which is discussed in \cite{fine-bh}.
 \begin{figure}[htbp]
   \centerline{\includegraphics[height=2cm]{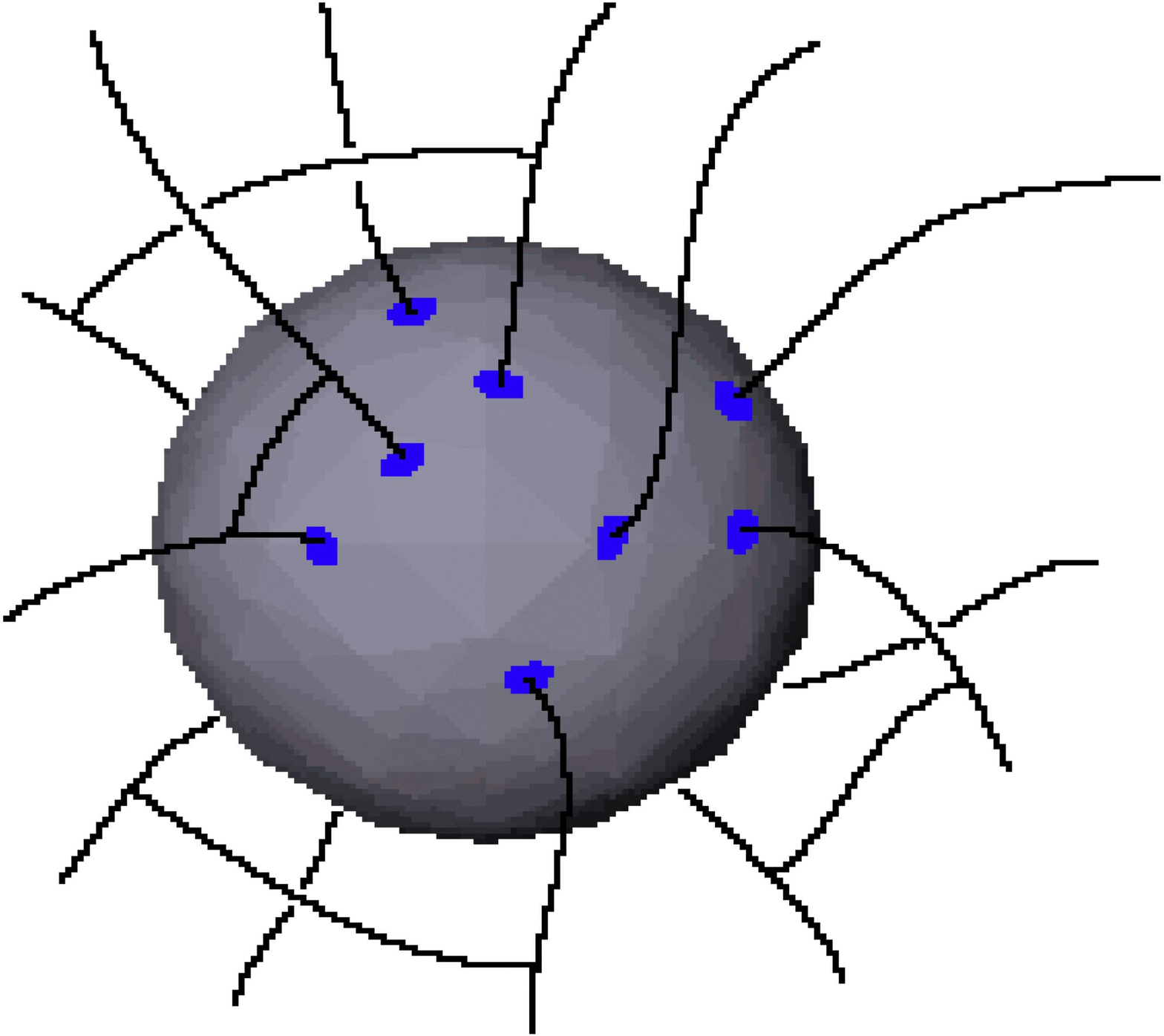}}
     \caption{A black hole in LQG.}
   \label{black_hole}
\end{figure}

\subsection{Heat and the cosmological constant}

There turns out to be a natural role for the cosmological constant, which is that it
parameterizes a quantum deformation of the algebra $\cal A$. For the case of
 $3+1$ dimensions, this leads to $\cal A$ being 
$SL_q (2)$ with $q=e^{\frac{2 \pi \imath}{k+2}}$ where the level $k$ is 
given by\cite{linking,qdef,cosconst}
\f
k = \frac{6\pi}{G \Lambda}   
\label{qdeform}
\ff

The quantum deformation of the symmetry algebra has a simple physical meaning, at least for $\Lambda >0$.  The ground state should be deSitter spacetime, which has an  horizon with an area 
\f
A= \frac{12 \pi}{\Lambda}
\ff
By the Bekenstein bound there should be a finite number of degrees of freedom observable on the horizon, given by 
\f
N= \frac{A}{4G\hbar}=  \frac{3\pi}{G\hbar \Lambda}
\label{Nbound}
\ff
This relationship has been called the $N-$bound and has been conjectured by Banks and Fishler to be fundamental\cite{Nbound}. 
If an observer rotates they see the horizon rotate around them, hence these degrees of freedom should fall into a single irreducible representation.  But if the Bekenstein bound is a real limit, there should not be any irreducible representation with more than $N$  states in it. This is precisely true if the rotational symmetry is quantum deformed by (\ref{qdeform}).   Thus the $N-$bound is a consequence of the quantum deformation of the symmetry induced by the cosmological constant\cite{cosconst}. 

A consequence of the quantum deformation of the label set is that the graphs are framed, so edges are represented by ribbons or tubes\cite{linking,qdef,knot-theory}.

A classic result of quantum field theory in curved spacetime is that $QFT$'s on the background of deSitter spacetime are thermal, with a temperature
\f
T= \frac{1}{2 \pi} \sqrt{\frac{\Lambda}{3}}
\label{TdS}
\ff
It turns out that one can extend this to quantum gravity at the non-perturbative level using a simple argument based on the few facts we have already mentioned.  The key is that deSitter spacetime corresponds to the solution of the topological field theory
(\ref{dS}.).  In terms of the configuration and momenta variables of the Ashtekar formulation, which are the $SU(2)_L$ connection $A_a^i$ and its conjugate momenta
$\tilde{E}^a_i \approx e \wedge e$ this becomes
\f
F^i_{ab} + \frac{\Lambda}{3} \epsilon_{abc} \tilde{E}^c_i =0
\ff
One can solve this with a Hamilton-Jacobi function on configuration 
space\cite{chopin-lee,cosconst}, which is a function $S(A^i)$
such that $\tilde{E}^a_i  = \frac{\delta S(A^i )}{\delta A_a^i}$.  This leads to the equation
\f
F^i_{ab} + \frac{\Lambda}{3} \epsilon_{abc} \frac{\delta S(A^i )}{\delta A_a^i}=0
\ff
There is also the Gauss's law constraint which requires that
\f
{\cal D}_a\tilde{E}^a_i= {\cal D}_a \frac{\delta S(A^i )}{\delta A_a^i} =0
\ff
These have a unique solution
\f
S(A^i) = -\frac{k}{4\pi} \int Y_{CS} (A^i )
\ff
where $Y(A^i )$ is the Chern-Simons invariant. Thus we can consider the Chern-Simons invariant to be a time functional on the Euclidean configuration space. 

If we choose  $\Sigma = S^3$ we find that there is a periodicity due to the property that under
large gauge transformations with winding number $n$
\f
\int Y_{CS} (A^i ) \rightarrow \int Y_{CS} (A^i ) + 8 \pi^2 n
\ff
This means that the Euclidean configuration space is a cylinder which further implies that {\it all correlation functions, for any fields in the theory} are periodic in an imaginary time variable given by the Chern-Simons functional $S(A^i )$.   But by the $KMS$ theorem, this means that the theory is at a finite temperature.  
If one works out the periodicity one finds precisely the temperature (\ref{TdS}).  

This applies to the full quantum gravity theory because it means that any quantum state on the full configuration space of the theory will be periodic in imaginary time.  Thus, with very little effort we greatly extend the significance of the deSitter temperature. This is an example of the power of seeing general
relativity in terms of connection variables and it is also an example of the importance of topological field theory to the physics of quantum gravity. 

\section{The problem of the emergence of classical spacetime}  

We have just seen that LQG gets several things about gravitational physics right, including the entropy of horizons and the temperature of deSitter spacetime. There are a number of other results that tell us that LQG and related theories have real physics we know in them. One thing that was done early in the development of the theory was to investigate classes of semiclassical states and show that their excitations, in the long wavelength limit were massless, spin two particles, i.e. gravitons\cite{weaves}.  It was further shown that when the theory was coupled to matter fields, one could recover the matter QFT on a classical background by expanding around semiclassical 
states\cite{chopin-lee}. 

This is encouraging, but  we should ask more. We want to show that these results follow  from expanding around the true ground state of the theory.  
As the fundamental hilbert space is described in combinatorial and algebraic terms, the key issue is that  classical spacetime is not fundamental, it must be an emergent, approximate description, analogous to  thermodynamics.  This was a problem that took some time to develop the tools to address, but in the last year there have been four separate developments that represent progress.

\begin{enumerate}

\item Rovelli and collaborators have computed the graviton propagator in spin foam models\cite{carlo-prop}.  They work in the Euclidean theory and fix a boundary, which is a four sphere, large in Planck units. They compute the amplitude for a graviton to travel from one point on the boundary to another, through the interior, which they treat by  a particular form of the spin foam path integral.  They get the right answer in the long wavelength limit. This shows that the theory has gravitons and reproduces Newton's gravitational force law. 

\item  Freidel and Livine have computed the spin foam path integral for $2+1$ gravity coupled to  matter\cite{etera-laurent}. They derive an effective field theory for the matter, by which they show that the full effect of quantum gravity in this case is to deform the symmetry of flat spacetime from the Poincare group to a quantum group called $\kappa$-Poincare\cite{DSR}. I will discuss the  meaning of this below. 

\item Ambjorn, Jurkiewicz and Loll have constructed a simple discrete and background
independent model of spacetime, which implements discreteness and causal structure, called the causal dynamical triangulations model\cite{CDT}. They find that it has a continuum limit which defines a theory which has a large universe limit. They can measure the dimension of spacetime by several means and it is to within error $3+1$.  

\item  Krebs and Markpoulou have proposed new criteria for the emergence of classical spacetime in terms of quantum information theory\cite{fotini-david}. 
They address the low energy physics by asking whether there are local excitations that remain coherent in spite of the fact that  they are continually in interaction with 
the quantum fluctuations in the geometry.  The answer is that excitations will remain 
coherent when  they are protected by
emergent symmetries.  The idea is then to analyze the low energy physics in terms 
of the symmetries that control the low energy coherent quantum states rather than 
in terms of emergent classical geometry.  To address this problem it was 
shown that one can apply the technology of  noiseless subsystems, or NS, 
from quantum information theory\cite{NS}. In this framework subsystems 
which propagate coherently are identified by their transforming under emergent
symmetries that commute with the interactions of the subsystem with an environment.
In this way they protect the subsystems from decoherence.   In the application of 
this idea to quantum gravity proposed in \cite{fotini-david}, the environment is 
the quantum fluctuations of geometry and the emergent particle states are to be 
identified as noiseless subsystems\cite{fotini-tomasz}.  

\end{enumerate}

\section{Possible new generic consequences}
 
Given that there is progress on this key issue, we can go on to discuss three more generic
consequences which might be associated with the low energy behavior of quantum theories of gravity. 
 
\subsection{Deformed special relativity}
 
A new physical theory should not just reproduce the old physics, it should lead to new predictions for doable experiments.  The problem of the classical limit is important not just to show that general relativity is reproduced, but to go beyond that and derive observable quantum gravity effects. It turns out that such effects are observable in quantum gravity, from experiments that probe the symmetry of spacetime.  

A big difference between a background independent and background dependent theory is that only in the former is the symmetry of the ground state a prediction of the theory.  In a theory based on a fixed background, the background, and hence its symmetry are inputs. But a background independent theory must predict the symmetry of the background.  

There are generally three possibilities for the outcome:

1) Unbroken Poincare invariance

2) Broken Poincare invariance, so there is a preferred frame\cite{broken}.

3) Deformed Poincare invariance or, as it is sometimes known, deformed or double special relativity (DSR)\cite{DSR}.

There is a general argument why the third outcome is to be expected from a background independent theory, so long as it has a classical limit. As the theory has no background structure it is unlikely to have a low energy limit with a preferred frame of reference. This is even more unlikely if the dynamics is instituted by a Hamiltonian constraint, which is essentially the statement that there is no preferred frame of reference.   Thus, we would expect the symmetry of the ground state to be Poincare invariance.  But at the same time, there is as we have described above, a discreteness scale, which is expected to be the minimal length at which a continuous geometry makes sense.   This conflicts with the lorentz transformations, according to which there cannot be a minimal length.  

The resolution of this apparent paradox is that the symmetry can be DSR, which is a deformation of Poincare invariance that preserves two invariant scales, a velocity and a length.  

There are then two questions.  Are there consistent interacting quantum theories with DSR symmetry? And if so, is DSR a generic prediction of background independent quantum gravity theories?  

The results mentioned above by Freidel and Livine show that DSR is the correct description for quantum gravity, coupled to matter in $2+1$ dimensional worlds\cite{etera-laurent}.  This answers the first question positively.  What about $3+1$?

There are heuristic calculations that indicate that $LQG$ in $3+1$ dimensions has 
a semiclassical approximation characterized by DSR\cite{falsifiable}. But there is as yet no rigorous 
proof of this.  One reason to expect a $DSR$ theory is to notice that the symmetry group of the ground state of the theory with a non-zero cosmological constant is, by 
(\ref{qdeform}), 
the quantum deformation of the deSitter or Anti-deSitter algebra. The contraction of this
is, under plausible assumptions for the scaling of the energy and momentum generators, no longer the Poincare algebra, it is the $\kappa$-Poincare algebra that characterizes
$DSR$ theories\cite{contract-DSR}.  

The three possibilities are distinguished by different experiments in progress.  We expect that a DSR theory will show itself by a) the presence of a $GZK$ threshold and similar thresholds for 
$Tev$ photons but b) a first order in $l_{Pl}$ and parity even  increase of the speed of light with
energy\cite{falsifiable}.   This is in contrast to the implications of breaking Lorentz invariance, which are a parity odd energy dependent shift in the speed of light and a possible shift in the $GZK$ threshold.

\subsection{Emergent matter}

In this and the next section I would like to describe two new possible generic consequences that have only recently been studied.

We are used to thinking that causal spin-network theories are theories of the quantum gravitational field alone. The problem of unification with fermions and the other forces is then postponed. This turns out to be  wrong.  In fact, it has recently been realized that many causal spin network theories have emergent local degrees of freedom that can be interpreted as elementary particles\cite{other,braids}.  
That this is a feature of loop quantum gravity and similar theories that might have been realized long ago, but it was only recently understood due to the application of the noiseless subsystem methods of Krebs and Markopoulou\cite{fotini-david}.  
The reason is that there are emergent quantum numbers which measure knotting and braiding of the embeddings of the graphs\cite{braids}. These are preserved under 
some forms of the local moves: no matter how many local moves are applied there are features of the braiding of edges which are conserved. 

These emergent conserved quantum numbers label local structures like braiding, which then can be seen to label noiseless subsystems of the quantum geometry.   
 \begin{figure}[htbp]
   \centerline{\includegraphics[height=2cm]{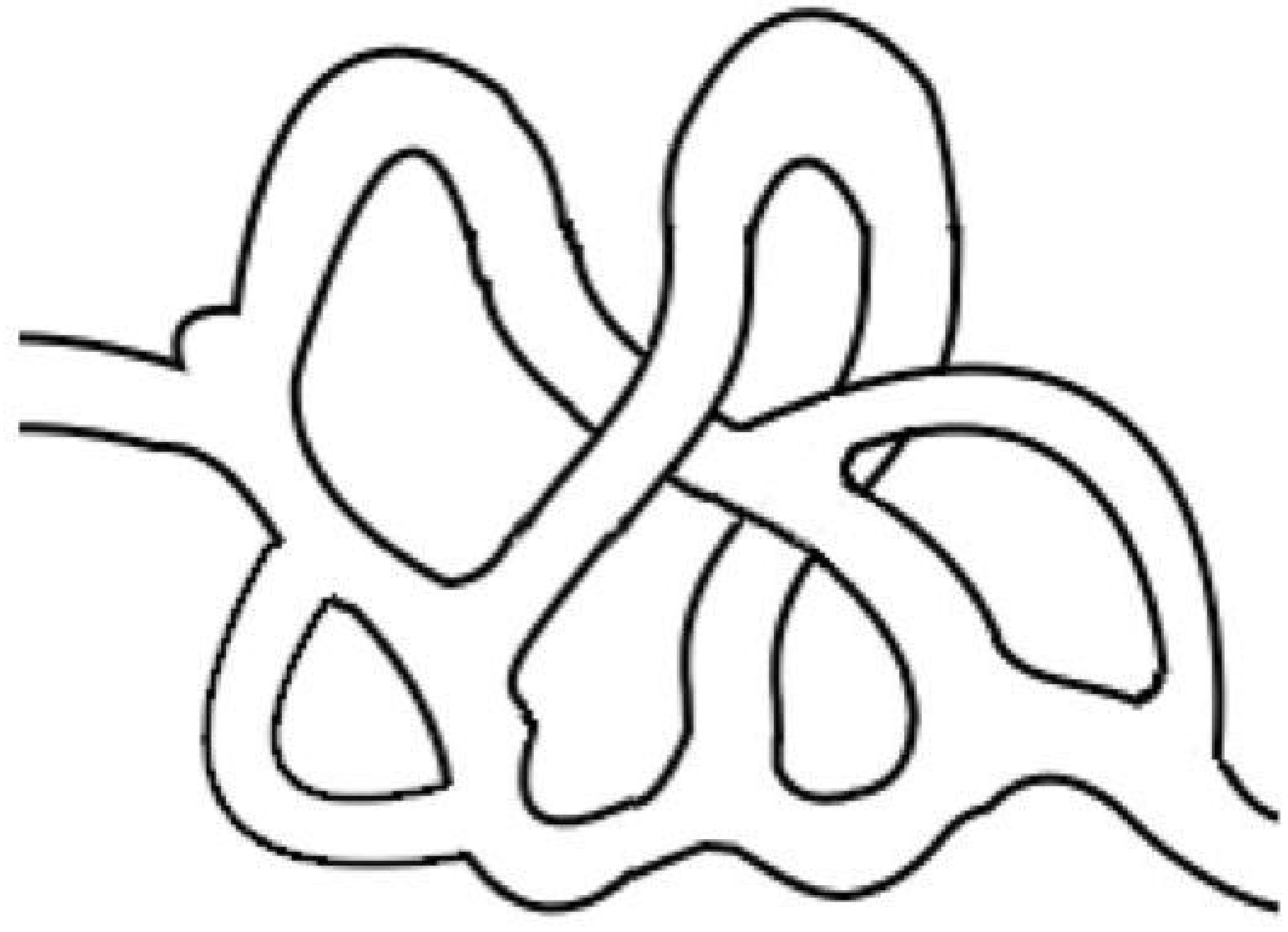}}
     \centerline{\includegraphics[height=2cm]{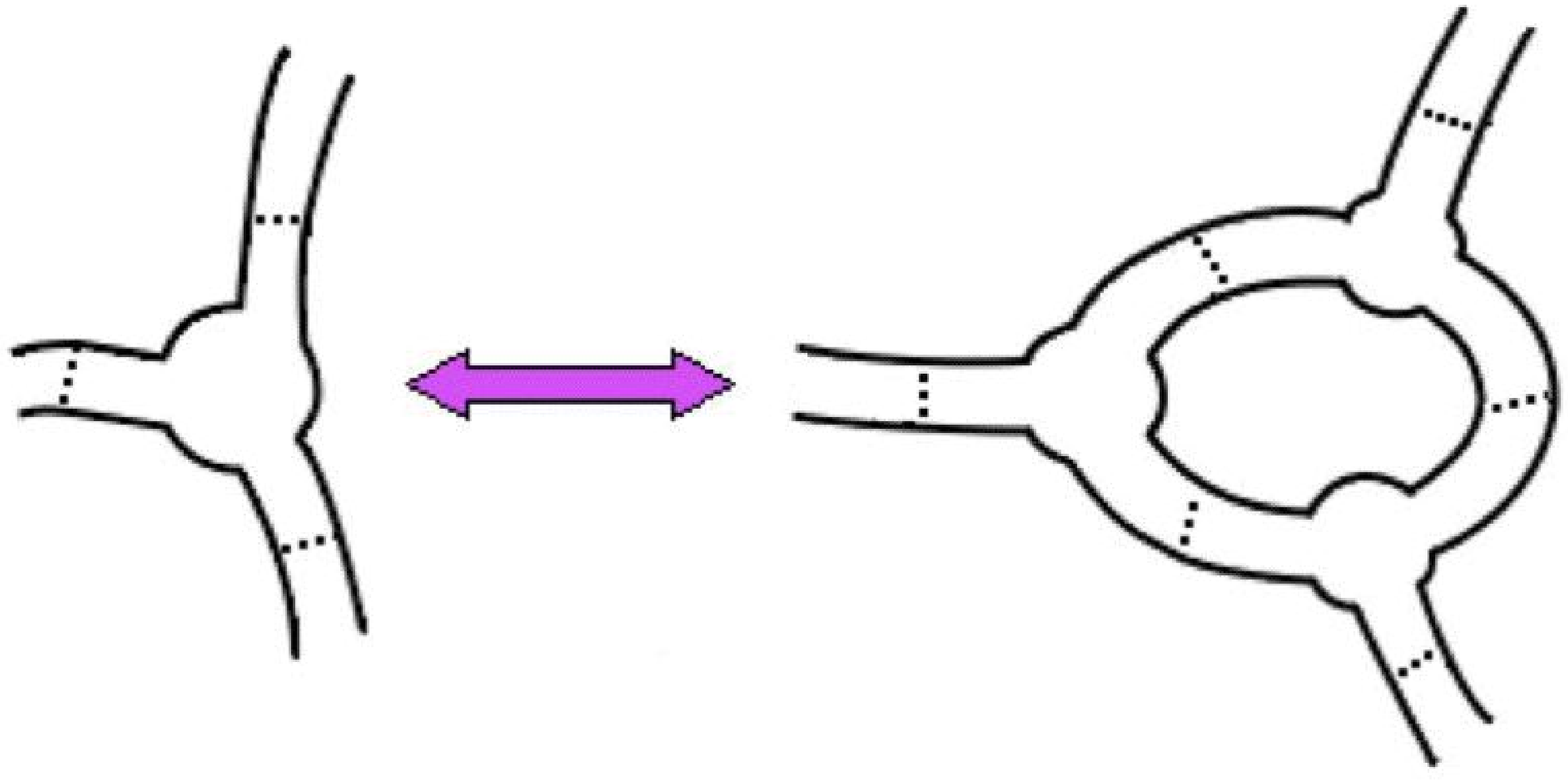}}
       \centerline{\includegraphics[height=2cm]{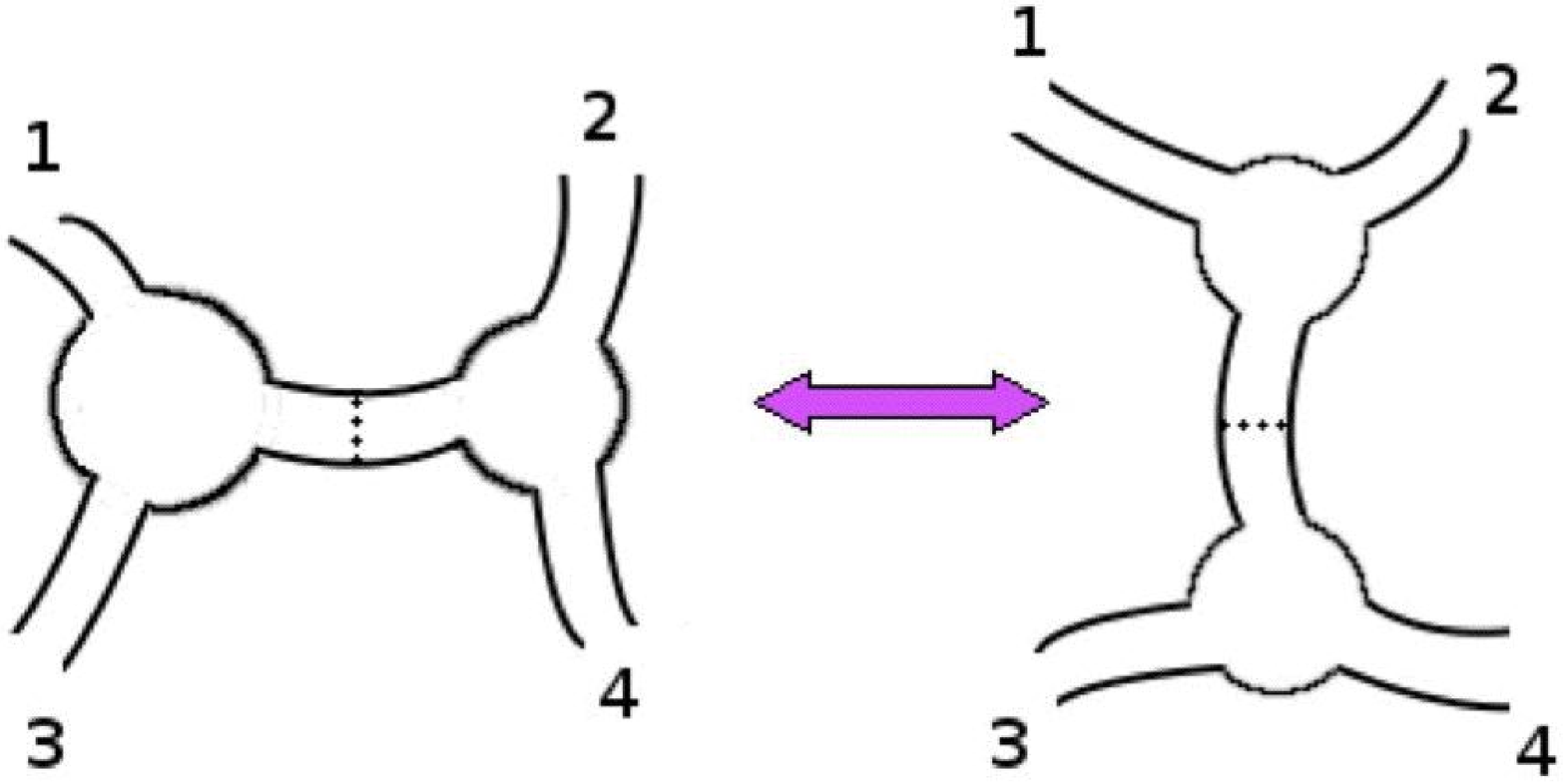}}
     \caption{Framed graphs or ribbons and their local moves.}
   \label{ribbons}
\end{figure}
 \begin{figure}[htbp]
   \centerline{\includegraphics[height=2cm]{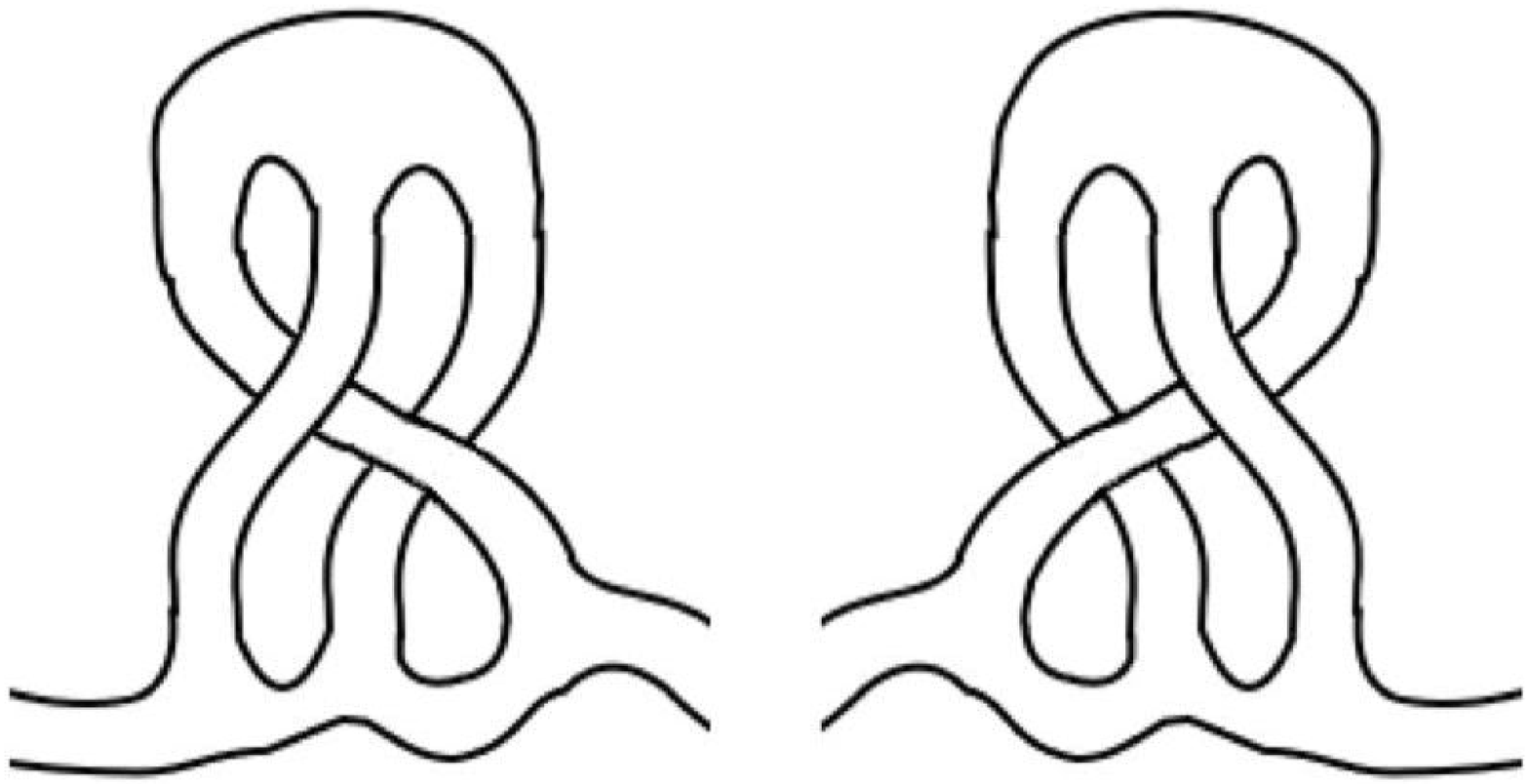}}
      \centerline{\includegraphics[height=2cm]{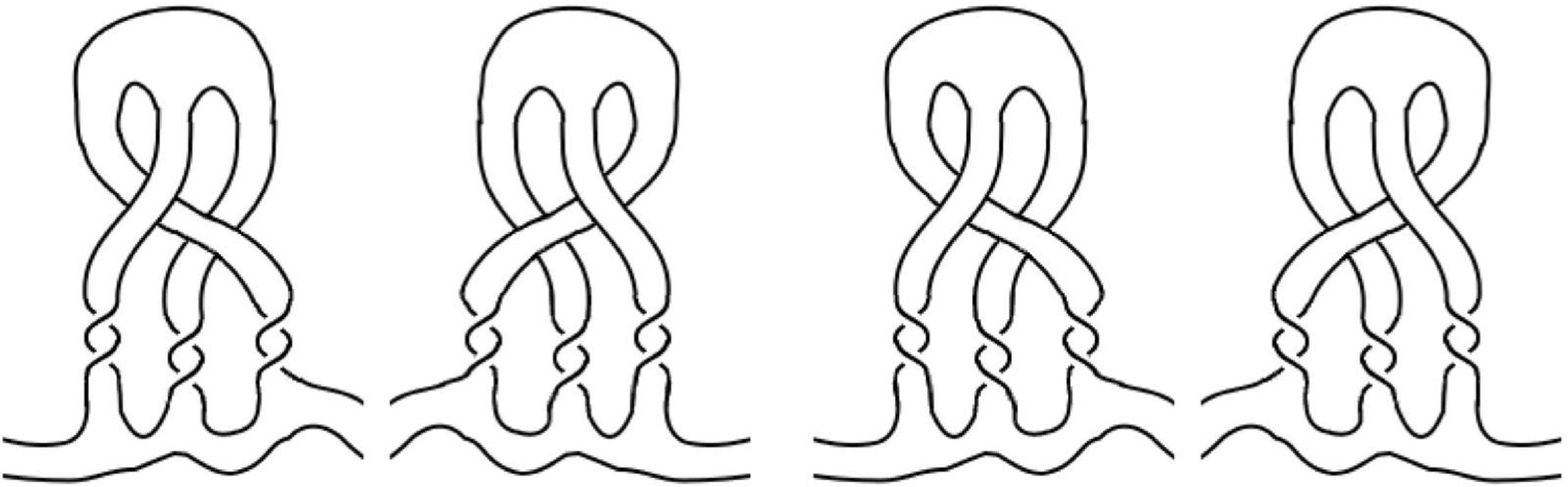}}
     \caption{Some braid state preserved under local moves. 
     Under the correspondence proposed in \cite{sundance,braids}. 
     the first set correspond to neutrinos, the second to electrons.}
   \label{fig:braidstates}
\end{figure}

One very interesting example of this is in theories with non-zero cosmological constant, in which case the relevant graphs are framed, and are represented by ribbons embedded in $\Sigma$. (See Figure~\ref{ribbons})  The simplest of these conserved local states, preserved
under the rules shown in turn out to correspond, with one additional assumption, to the first generation of quarks and leptons of the standard model\cite{sundance,braids}. Some of these are shown in 
Figure~\ref{fig:braidstates}

We see from this that causal spin network theories including loop quantum gravity are also unified theories, in which matter degrees of freedom are automatically included. It is also very interesting that the classification of these emergent matter degrees of freedom appears to depend only weakly on the properties of the theory, and so are generic over large classes of theories. 
 
\subsection{Disordered locality}

Each spin network state, $\Gamma$ has a {\it microscopic locality} given by the connections in the graph. 
Let us suppose that a semiclassical state exists,
\f
|\Psi > = \sum_\Gamma a_\Gamma |\Gamma > 
\label{macro}
\ff
corresponding to a classical
spatial metric $q_{ij}$.   That metric defines a notion of {\it macroscopic locality}.  The correspondence may  be
defined by measurements of coarse grained geometrical observables, such as volumes and areas.  We may
also require that excitations of $|\Psi >$, corresponding to graviton or matter degrees of freedom, propagate as if
they were on the background metric $q_{ij}$.   

But, as argued by Markopoulou, in \cite{f-this}, it may still not be the case that there is a complete
correspondence between the macrolocality defined by  $q_{ij}$ and the microlocality defined by
some or all of the graphs $\Gamma $ whose states have significant amplitude in (\ref{macro}).   

Consider, for example, the case of a ``weave state", which is a random lattice constructed to approximate a flat background metric $q^0_{ij}$ on a torus $T^3$.  This consists of
a graph $\Gamma_0$ embedded in the torus  such that only nodes of order Planck distance apart in $q^0_{ab}$ are connected. The spins and labels on nodes are chosen so that measurements of areas and volumes in the state $|\Gamma_0>$ coincide with the metric $q^0_{ij}$. Let the total volume be $V=N l_{Pl}^3$ for some
very large $N$.  We can then, for example, choose $\Gamma_0$ to be four valent with $N$ nodes and
$2N$ edges.  Such a $|\Gamma_0>$ is a state corresponding to the metric $q^0_{ij}$ in which microlocality
and macrolocality coincide. 

But now let us  add to the graph $\Gamma_0$, a new link connecting two randomly chosen nodes
of $\Gamma_0$. It is not hard to see that we can adjust the labels on the edges and nodes so that 
no large areas or volumes are changed.  In fact, we can do this $M$ times, at least so long
as  $M<<2N $, without changing any large areas or volumes. 
Each of these $M$ new links  connects two randomly chosen nodes of
$\Gamma_0$, making a new graph $\Gamma^\prime$.   The corresponding  state 
$|\Gamma^\prime >$ is still a
semiclassical state for the metric $q^0_{ab}$ and will reproduce it when sufficiently coarse
grained observables are measured.  But it provides an example of Markopoulou's observation
that micro and macro notions of locality need not coincide even in the low
energy limit\cite{nonlocal}.  We may call this phenomena, {\it disordered locality.}   

At first sight it seems as if disordered locality would kill the theory, because there would be macroscopic violations of locality in the low energy limit.  But it turns out that this need not the case, if the disagreement between micro and macro locality is rare enough.  
For example,  suppose that the probability that a node has a non-local edge, 
$p = \frac{M}{2N}$ is on the order of $10^{-100}$.  This would still mean there are on the order of
$10^{80}$ random non-local edges within the Hubble volume.  Could we do any measurements to tell that the quantum geometry of our universe was based on
$\Gamma^\prime$ rather than $\Gamma$?  
 \begin{figure}[htbp]
   \centerline{\includegraphics[height=2cm]{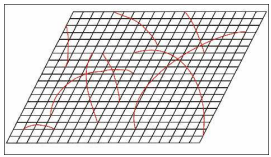}}
     \caption{A lattice with disordered locality from a contamination of non-local links.}
   \label{baidstates}
\end{figure}

It would be very unlikely that any two nodes within the earth are connected by one of the
non-local edges, so there would be very hard to directly detect non-locality. Moreover, since the defects were at the Planck scale, the amplitude for low energy quanta  to jump across a non-local link would be 
suppressed by $l_{Pl}^2 E^2$.  So we are unlikely to see fermions appearing and disappearing across the links.    Moreover since the whole universe is in thermal equilibrium at the same temperature the transfers of energy through the non-local links would also be hard to observe. Studies have been done of the thermodynamics of spin systems on networks with disordered locality and the main effect for small $p$ is to raise the Curie temperature by an amount of order $p$ without strongly affecting the correlation functions\cite{eaton}.  

Would dynamics suppress such non-local links? The answer is that the dynamics cannot. The reason is that the dynamics is micro-local, 
and hence defined by the connectivity of the graph $\Gamma^\prime$.  The local moves that generate the dynamics cannot remove non-local edges connecting two nodes far away in 
$\Gamma^0$ and hence $q^0_{ab}$.  This was shown for stochastic evolution of graphs
in \cite{hal}, but there is no reason to believe the results will be different for quantum evolution. 

Where then would the effects of disordered locality show up?  
The following are speculative suggestions which are presently under investigation.

\begin{itemize}

\item{}At cosmological scales there may be new effects coming from the fact that the shortest distance between points will go through the non-local edges.  Could this have something to do with the dark energy and dark matter problems\cite{withisabeau}?

\item{}If electric flux is trapped in a non-local edge its ends look like charged particles. 
This provides a quantum mechanical version of Wheeler's old hypothesis that matter
comes from charged wormholes.

\item{}Suppose we have a subsystem, large enough to contain the ends of many non-local links but small enough that almost all of these connect it to the rest of the universe. 
Even at zero temperature the subsystem is subject to a random disorder coming from its connections to the rest of the universe through the non-local links. There are results that indicate that this could be the origin of quantum phenomena\cite{linkquantum}  

\item{}Non-local links could connect regions of the universe to others beyond the horizon.  This could provide a solution to the horizon problem without inflation. Could it also lead to the generation of a scale invariant spectrum of fluctuations? This is  
discussed next\cite{CMB}.  

\end{itemize}

\subsection{Disordered locality and the CMB spectrum}

Here is a simple estimate that shows that effects of disordered locality could be 
responsible for the power spectrum observed in the $CMB$\cite{CMB}.  
Assume that there is a random (and hence scale invariant) distribution of pairs of 
points in the universe that are connected by a non-local link. We call these pairs
$x_i$ and $y_i$, for $i= 1,...,N_{NL}$.  For practical purposes these pairs can be considered to be identified, as they are the equivalent of a Planck distance apart.   We can estimate the contribution these points make to the two point correlation function for energy
fluctuations, as 
\f
D(x,y)_{NL} = < \frac{\delta \rho }{\rho} (x)\ \frac{\delta \rho }{\rho} (y) > =
l_{Pl}^2 T^2 \sigma^2_U    \sum_{i =1}^N \delta^3 (x,x_i) \delta^3 (y,y_i) 
\ff 
The factor $l_{Pl}^2 T^2 $ is due to the cross section of a planck scale edge being roughly the Planck
area.  The factor 
$\sigma^2_U$ is the local (because the points connected by a non-local link are identified) fluctuation in energy.
\f
\sigma^2_U = \frac{<E^2> - <E>^2}{<E>^2} = \frac{T}{\rho V}= \frac{1}{VT^3}
\ff
were $T$ is the temperature and $V$ is the volume of space within the horizon.
The power spectrum is related to the Fourier transform
\f
D(k)_{NL} = \int_V d^3 x \int_V  d^3y D(x,y)_{NL}  e^{\imath k\cdot (x-y)}
\ff
Since the connected pairs are distributed randomly, we find the correct scale invariant
spectrum of fluctuations, 
\f
D(k)_{NL}= \frac{A}{V k^3} \ .
\ff
This should hold outside the horizon at decoupling, when there are no other long ranged correlations possible.  
The amplitude is given by
\f
A= 2\pi^2 l_{Pl}^2 T^2  N_{NL} \sigma^2_U 
\ff
If we evaluate $\sigma^2_U$ at decoupling we find around $10^{-90}$. This tells us
that we get the correct amplitude of $10^{-10}$ with an $N_{NL} \approx 10^{124}$.  
This gives us a $p \sim 10^{-56}$ which from the above discussion is well within observable limits. 
This is very rough, but it shows that distributed locality can 
comfortably do the job inflation does of solving the horizon problem in a way that leads to a scale invariant distribution of fluctuations outside the horizon, of the observed amplitude. 

 \section{Conclusions}
 
To summarize, the causal spin network theories, including loop quantum gravity and spin foam models,  do a number of things that are expected of
any sensible quantum theory of spacetime.  They are 
finite, they predict that quantum geometry is discrete, they remove spacelike singularities and explain the entropy of black hole and cosmological horizons as well as the temperature of deSitter spacetime.  
If one adds to this that there is progress understanding whether and how classical spacetime emerges from the quantum geometry,  we see that these continue to show promise as plausible models of quantum gravity.  While there is certainly still much to do, the last years have given us a well defined foundation to build on.  

But theories triumph not because they do what  is expected, but because of the surprises they lead to. A good theory must predict new phenomena, which are then observed.
In the case of causal spin network theories we see several unexpected consequences which all have implications for experiment and observations.  These are
\begin{itemize}

\item{}The symmetry of the ground state is DSR, leading to an energy dependent, parity even,  speed of light.


\item{}There is evidence that $LQG$ predicts that spacelike singularities bounce. This opens up the possibility of tuning the parameters that govern low energy physics through a dynamnical mechanism like cosmological natural selection (CNS)\cite{CNS}.  

\item{}These theories have emergent local degrees of freedom, hence  they automatically unify geometry and matter.  

\item{}Disordered locality has consequences for cosmological observations because
even at small levels that make it unobservable in local experiments it dominates in the early universe and at cosmological scales.  A rough estimate of such effects shows that this mechanism has a possibility to naturally solve the horizon problem while predicting the correct spectrum of fluctuations of the $CMB$.  

\end{itemize}

So which kind of theories are $LQG$ and other causal spin network theories?  Are they the good kind of unification that leads to consequences we celebrate or the embaressing kind that lead to consequences that must be hidden.  
The discovery that these theories generically predict emergent particle states certainly leaves them vulnerable to
quick falsification. While there is preliminary evidence that a large class of theories can reproduce some features of the standard model, there is a lot that these theories  have to get right so as not to disagree with observation.   

Disordered locality certainly offers other possibilities for falsification.  If the deviations from locality are small, disordered locality gives rise to new mechanisms for solving hard problems like the horizon problem and dark energy.  This means they lead to falsifiable predictions, for there is only one parameter, $p$ which controls these effects.  But what if the deviations from locality are not small? One possibility is the proposal of Markopoulou,
who argues  in \cite{f-this} that the macroscopic causal structure will be defined by the interactions of the coherent excitations which are the elementary particles.  As described there, the test of this program is then whether the Einstein equations are reproduced.  

  Finally, the expectation the the low energy limit is $DSR$ has to be counted as fortunate, as this experiments sensitive enough to test the implied predictions are expected in the next few years. 

Thus, there appears to be a good possibility to use these generic consequences to test whether the correct unification of spacetime and quantum theory is in terms of a causal spin network theory.  In the next few years we may hope to sharpen up the arguments described here to detailed predictions that may be confirmed or falsified in upcoming 
experiments.

\section*{Acknowledgements}

 I would like to thank the many people who contributed the ideas and results described here, for many discussions over the years. The proposal that locality is disordered in background independent theories, and the observation that these theories have emergent particle states, are due to Fotini Markopoulou.  I would also like to thank
M Ansari,  S. Bilson-Thompson, O Dreyer, H. Finkel, T. Konopka,  J. Magueijo, S. Majid, J. Moffat,   M. Paczuski, I Premont-Schwarz and Y.  Wan 
for collaborations and discussions which were very helpful for exploring these new ideas. 



\end{document}